\documentclass[a4paper,12pt,oneside]{article}

\usepackage{amsmath,amssymb,graphicx}

\def\om{\omega}

\def\lb{\left[}

\def\be{\begin{equation}}
\def\ee{\end{equation}}
\def\ai{\'{\i}}
\def\ba{\begin{array}}
\def\ea{\end{array}}
\def\bea{\begin{eqnarray}}
\def\eea{\end{eqnarray}}
\def\in{\infty}
\def\t{\theta}
\def\lb{\label}
\def\be{\begin{equation}}
\def\ee{\end{equation}}
\def\d{\delta}
\def\f{\phi}
\def\pd{\partial}

\usepackage{mathrsfs}
\usepackage{amsmath}
\usepackage{amssymb}
\usepackage{graphicx}
\usepackage{tabularx}
\usepackage[latin1]{inputenc}
\usepackage{enumerate}
\usepackage{graphicx,amssymb,amstext,amsmath}
\usepackage[bottom=3.5cm, textheight=22.5cm, textwidth=15cm]{geometry}
\usepackage{cancel}
\usepackage[para]{footmisc}

\begin{document}

\baselineskip0.25in
\title{Probing  global aspects of a geometry by the self-force on a charge: cylindrical thin-shell wormholes}
\author{E. Rub\ai n de Celis$^{1,3}$\footnote{e-mail: erdec@df.uba.ar}\,, O. P. Santillan$^{2}$\footnote{e-mail: osantil@dm.uba.ar}\,$\,$ and C. Simeone$^{1,3}$\footnote{e-mail: csimeone@df.uba.ar}\\
{\footnotesize   $^1$ Departamento de F\ai sica, Facultad de Ciencias Exactas y Naturales,} 
\\
{\footnotesize Universidad de Buenos Aires, Ciudad Universitaria Pab. I, 1428,Buenos Aires, Argentina.}
\\
{\footnotesize $^2$ CONICET--Instituto de Investigaciones Matem\'aticas Luis Santal\'o,}
\\
{\footnotesize Ciudad Universitaria Pab. I, 1428, Buenos Aires, Argentina.}
\\
{\footnotesize $^3$ IFIBA--CONICET, Ciudad Universitaria Pab. I, 1428, Buenos Aires, Argentina.}}
\maketitle

\begin{abstract}
We obtain the self-interaction for a point charge in the space-time of a cylindrical thin-shell wormhole connecting two identical locally flat geometries with a constant deficit angle. Although this wormhole geometry is locally indistinguishable from a cosmic string background, the corresponding self-forces are different even at the quali\-tative level. In fact in the cosmic string geometry the force is always repulsive while for the wormhole background  we find that the force may point outwards or towards the wormhole throat depending on the parameters of the configuration. These results suggest that the study of the electromagnetic fields of charged particles is a useful tool for testing the global properties of a given background.
\end{abstract}

\section{Introduction}

 Wormholes are space-time geometries with a nontrivial topology, so that two regions of the universe are connected by a traversable throat; the throat is a  surface where the geodesics open up. Following the first work by Morris and Thorne \cite{mo}, traversable wormholes have received considerable attention \cite{visser}, because they would imply some notable consequences, as for example the possibility of time travel \cite{mo-fro}. However, the necessity of exotic matter (i.e. matter not fulfilling the energy conditions), which seems unavoidable within the framework of general relativity, is a central objection against the actual existence of wormholes. This point has been carefully studied for wormholes of the thin-shell class \cite{povi,visser},  which are mathematically constructed by cutting and pasting two manifolds, so that exotic matter is restricted to a layer located at the throat.

The geometrical aspects of topological defects as cosmic strings have been the object of a detailed study in the last three decades, mainly because they could have played an important role in structure formation in the early universe; it was also noted that they could be observed by their gravitational lensing effects (see Ref. \cite{vilenkin}). On the other hand, strings are the objects more seriously considered by present theoretical developments as the fundamental building blocks of nature.  Therefore, the interest in the gravitational effects of both fundamental and cosmic strings has been  renewed in the last years (for example, see Refs. \cite{strings}). As a natural consequence, cylindrically symmetric wormhole geometries, as those associated to cosmic strings, have been recently considered  \cite{cil}. Thin-shell wormhole configurations associated to local and global cosmic strings have been studied in Refs. \cite{cilts1,cilts2}. In particular, the case of a cylindrical locally flat wormhole geometry with a deficit angle, associated to a straight gauge string, was treated in \cite{cilts1}.

For any thin-shell wormhole, beyond the throat the geometry is locally indistinguishable from the metric from which the mathematical construction starts. For example, the space-time of the cylindrical wormhole studied in \cite{cilts1} is locally the same of a cosmic string metric, and only the possibility to probe the global properties of the geometry would allow an observer to decide between both topologically different manifolds. Our proposal is that, for any thin-shell wormhole, such a test of the global behaviour of the geometry can be provided by the  electrostatics of such a simple system as a point charge: A gravitational field changes the electric field of a point charge so that a finite self-force appears. Even though the equivalence principle states that, for a freely falling observer, near a point charge the Maxwell equations of  flat  space-time hold, this does not mean that the solution of these equations must be locally the same for flat space-time: such a solution would not fulfill the correct boundary conditions in the infinity, that is, it would not have the asymptotic behaviour which results from the space-time curvature. Hence, the solution which is asymptotically well behaved is locally different from that of flat space-time. In particular, the field is no more symmetric around the charge, and this leads to the existence of a self-force. Now, the field is determined by the whole background geometry; hence the topology, that is the existence or not of a wormhole throat, must be revealed by the associated electrostatic self-force.

The calculation of the self-interaction starts from the Maxwell equations for curved space-time \cite{ll} (with $c=1$):
\begin{eqnarray}
\left(F^{\mu\nu}\sqrt{-g}\right)_{,\nu} & = & 4\pi j^\mu \sqrt{-g}\nonumber\\
F_{\mu\nu} & = & A_{\nu,\mu}-A_{\mu,\nu}\nonumber\\
j^\mu & = & \sum_a\frac{q_a}{\sqrt{-g}}\delta({\bf x}-{\bf x}_a)\frac{dx^\mu}{dx^0}\,,
\end{eqnarray}
where $\mu=(0,1,2,3)$, $g$ is the determinant of the metric, $A_\mu$ is the four-potential, $j_\mu$ is the four-current and $q_a$ are the associated  charges.
For only one rest charge $q$ located at the point ${\bf x}'$ we simply have
\begin{equation}
 \left(F^{0k}\sqrt{-g}\right)_{,k}=4\pi q\delta({\bf x}-{\bf x}') \, ,
\end{equation}
where $k=(1,2,3)$. The first case considered of a self-force of physical interest was a charge in a Schwarzschild metric; it was shown that the self-force on a  point charge $q$  is repulsive and has the form
\be
f\sim\frac{Mq^2}{r^3} \, ,
\ee
where $M$ is the mass of the spherical source and $r$ is the usual Schwarzschild radial coordinate of the probe charge. This result was first obtained within linearized general relativity \cite{vil79}, and was later confirmed  working in the framework of the full theory; see \cite{will80} and also \cite{varios}. After these works, the study of the self-interaction of a charge was then extended to many geometries. In particular, the self-force on a charge in the locally flat geometry of a straight cosmic string was found by Linet \cite{lin}; the force points outwards, and has the form
\be
f\sim\frac{\mu q^2}{r^2} \, ,
\ee
where $\mu$ is the mass per unit length of the string and $r$ is the distance from the string to the charge. This nonvanishing force, which is associated to the deficit angle, illustrates how the global properties of a manifold are manifested by electrostatics. The problem of a charge in the non trivial topology of wormhole geometries was recently addressed in a set of works \cite{whs} which in general agree in the interesting result of an attractive force, i.e., a force towards the wormhole throat. Here we will consider a cylindrical thin-shell wormhole which connects two identical locally flat geometries with a constant deficit angle. As pointed out above, this space-time is locally indistinguishable from the geometry around  a gauge cosmic string. We will obtain the self-interaction for a point charge in such a wormhole geometry, thus providing a tool to  probe the space-time beyond its local properties; the result would allow an observer to decide whether the topology is trivial, as it would be in the case of a cosmic string, or it includes a throat, thus showing the existence of a cylindrical thin-shell wormhole.

\section{Evaluation of the self-force}

The calculation of the self-force on a charge in the space-time of a cylindrical thin-shell wormhole connecting two cosmic string geometries is best understood by proceeding in successive steps: we first illustrate how to obtain the electrostatic potential in Minkowski space-time working in cylindrical coordinates; then we consider the cosmic string case, and finally we study the charge in the wormhole geometry. This allows to easily separate the finite and divergent parts of the potential, which is a central point of the evaluation of the self-interaction, and to clearly understand the difference between the cosmic string  and the wormhole results.

\subsection{Charge in Minkowski space-time}

Minkowski space-time in cylindrical coordinates $(t,r,\t,z)$ is described by the line element
\be
ds_M^2=-dt^2+dr^2+r^2d\t^2+dz^2  \, ,
\ee
where the ranges of the coordinates are:
$$-\in<t<+\in\,, \quad r\geq0\,, \quad0\leq\t\leq2\pi\,,\quad -\in<z<+\in \,.$$
The Maxwell equations for a static point charge located at  $(r',\t',z')$ in Minkowski space-time give
\be
\Delta A^0=-\frac{4 \pi q}{r}\,\d(r-r')\d(\t-\t')\d(z-z')
\ee
for the zero component $A_0=V_M$ of the four-potential, where $\Delta\equiv g^{kl}\nabla_k\nabla_l$, while the other components of the four-potential vanish. Hence in the Minkowski background we have the Poisson  equation
\be\lb{mm}
\Delta V_M = \left[\frac{\pd^2}{\pd r^2}+\frac{1}{r}\frac{\pd}{\pd r}+\frac{1}{r^2}\frac{\pd^2}{\pd \t^2}+\frac{\pd^2}{\pd z^2}\right] V_M= -\frac{4 \pi q}{r}\,\d(r-r')\d(\t-\t')\d(z-z') \, .
\ee
The solution of Eq. (\ref{mm}) can be expanded in terms of a complete set of orthogonal functions of the coordinates $\t$ and $z$ as follows:
\be\lb{vm}
V_M=q\int \limits_{0}^{+\in}dk\cos[k(z-z')]\sum_{n=0}^{+\in}a_n\cos[n(\t-\t')]g_{n}(r,r')  \, ,
\ee
where $n \, \epsilon \, \mathbb{N}_0$, and
\be
a_n = \left\{ \ba{ll}
         \frac{2}{\pi} & \mbox{if $n=0$} \,,\\
         \frac{4}{\pi} & \mbox{if $n>0$}\,.
         \ea \right.
\ee
Then $g_n(r,r')$ must satisfy the nth order inhomogeneous modified Bessel equation,
\be\lb{bessel}
\left[\frac{\pd^2}{\pd r^2}+\frac{1}{r}\frac{\pd}{\pd r}-\left(k^2+\frac{n^2}{r^2}\right)\right]g_n(r,r')=
-\frac{\d(r-r')}{r}\,,
\ee
with the requirement to be finite at $r=0$ and vanish at $r\rightarrow\infty$, i.e.
\be\lb{gn}
g_n(r,r')=
\left\{ \ba{ll}
         K_n(kr')I_n(kr) & \mbox{if $r\leq r'$} \,,\\
         K_n(kr)I_n(kr') & \mbox{if $r\geq r'$}\,.
         \ea \right.
\ee
In (\ref{gn}), $K_n(kr)$ and $I_n(kr)$ are the usual modified Bessel functions, two independent solutions of the homogeneous
version of Eq. (\ref{bessel}).

\subsection{Cosmic String}

The gauge cosmic string space-time in cylindrical coordinates $(t,r,\phi,z)$ has the metric \cite{vilenkin,vilgothis}
\be
ds_{CS}^2=-dt^2+dr^2+\om_0^2r^2d\phi^2+dz^2 \, ,
\ee
where the constant $\om_0$, such that $0<\om_0\leq1$, is determined by the mass $\mu$ per unit length of the string; as is common, we neglect the core radius and consider a string associated to an energy-momentum tensor $T_\mu^\nu=\rm{diag}(\mu,0,0,\mu)\,\delta(x)\,\delta(y)$, so that the ranges of the coordinates are
 $$-\in<t<+\in\,, \quad r>0\,, \quad0\leq\f\leq2\pi\,,\quad -\in<z<+\in \, .$$
The Poisson equation for the electrostatic potential $V_{CS}$ associated to a point-like charge $q$ located at $(r',\f',z')$ in the cosmic string manifold is
\be\lb{mcs}
\Delta_{(\om_0)} V_{CS} =
\left[\frac{\pd^2}{\pd r^2}+\frac{1}{r}\frac{\pd}{\pd r}+\frac{1}{\om_0^2r^2}\frac{\pd^2}{\pd \f^2}+\frac{\pd^2}{\pd z^2}\right] V_{CS}=
-\frac{4 \pi q}{\om_0r}\,\d(r-r')\d(\phi-\phi')\d(z-z').
\ee
Changing to a coordinate $\t\equiv\om_0 \f$, the equation for the potential  becomes
\be \lb{mcsm}
\Delta_{(\om_0)} V_{CS} = \left[\frac{\pd^2}{\pd r^2}+\frac{1}{r}\frac{\pd}{\pd r}+\frac{1}{r^2}\frac{\pd^2}{\pd \t^2}+\frac{\pd^2}{\pd z^2}\right] V_{CS}= -\frac{4 \pi q}{r}\,\d(r-r')\d(\t-\t')\d(z-z')\;,
\ee
which is the usual equation in the  space-time covered by the coordinate system $(t,r,\t,z)$ with $0\leq\t\leq2\pi\om_0$, and for a point charge located at $r=r'$, $\t'=\om_0 \f'$, and $z=z'$.

The electrostatic potential must satisfy:
\be
 \ba{ll}
V_{CS}(r,z,\t=0)=V_{CS}(r,z,\t=2\pi\om_0)\,,\\
\frac{\pd V_{CS}}{\pd\t}(r,z,\t=0) = \frac{\pd V_{CS}}{\pd\t}(r,z,\t=2\pi\om_0) \, .
         \ea
\ee
The solution of Eq. (\ref{mcsm}) can be expanded in a complete set of orthogonal functions of the coordinates $\t$ and $z$ as
\be\lb{vcs}
V_{CS}=\frac{q}{\om_0}\int \limits_{0}^{+\in}dk\cos[k(z-z')]\sum_{n=0}^{+\in}a_n \cos[\nu(\t-\t')]g_{\nu}(r,r') \, ,
\ee
where $\nu=n/\om_0$, $n \, \epsilon \, \mathbb{N}_0$ and
\be
a_n = \left\{ \ba{ll}
         \frac{2}{\pi} & \mbox{if $n=0$} \,,\\
         \frac{4}{\pi} & \mbox{if $n>0$}\, .
         \ea \right.
\ee
So for the cosmic string space-time, the radial part of the expansion of the potential is given by
\be
g_{\nu}(r,r')=
\left\{ \ba{ll}
         K_{\nu}(kr')I_{\nu}(kr) & \mbox{if $r\leq r'$} \,,\\
         K_{\nu}(kr)I_{\nu}(kr') & \mbox{if $r\geq r'$}\, ,
         \ea \right.
\ee
where now $K_{\nu}(kr)$ and $I_{\nu}(kr)$ are the modified Bessel functions of order $\nu=n/\om_0$.

In the neighborhood of the point charge, the electrostatic potential can always be written as
\be\lb{vcsh}
V_{CS}(r,\f,z)=V_M(r,\f,z)+H_{CS}(r,\f,z).
\ee
$H_{CS}$ is solution of the homogeneous equation (\ref{mcs}) and represents the nondivergent part of the electrostatic potential at the position of the charge. So the electrostatic energy of the point charge can be evaluated as:
\be
W_{CS}=\frac{1}{2}\,q\,H_{CS}(r',\f',z') \, .
\ee
Using the results from Eqs. (\ref{vm}) and (\ref{vcs}), the self-energy of the charge
$q$ for any position with $r>0$ is
\be\lb{ecs}
W_{CS}(r)=\frac{\,q^2}{2} \int \limits_{0}^{+\in}dk\sum_{n=0}^{+\in}a_n \left[\frac{1}{\om_0}\,K_{\nu}(kr)I_{\nu}(kr)-K_{n}(kr)I_{n}(kr)\right] .
\ee
A closed expression for this result was previously obtained by Linet \cite{lin}; in the Appendix it is checked to coincide for particular cases. The electrostatic self-force is given by
\be\lb{fcs}
f_{CS}^r(r)=-\frac{\pd}{\pd r}W_{CS}(r)\quad \textrm{and}\; f_{CS}^{\f}=f_{CS}^{z}=0\,.
\ee
The force turns out to be directed away from the cosmic string, and is proportional to $1/r^2$.

\subsection{Cylindrical thin-shell wormhole}

We now consider the case of a charge in the space-time of a thin-shell wormhole connecting two identical cosmic string geometries $\mathcal{M}_+$ and $\mathcal{M}_-$, that is, two locally flat geometries with the same deficit angle and a hole of radius $a$, which is the radius of the cylindrical wormhole throat \footnote{This configuration is obtained by placing at $r = a$ a shell of negative energy density, so the wormhole geometry is supported by exotic matter.}. In cylindrical coordinates $(t,r_{\pm},\phi,z)$ the metric reads  \cite{cilts1}
\be
ds_{W}^2=-dt^2+dr_{\pm}^2+\om_0^2r_{\pm}^2d\phi^2+dz^2,
\ee
with $0<\om_0\leq1$ and the coordinate ranges:
$$-\in<t<+\in\,, \quad r_{\pm}\geq a\,, \quad0\leq\f\leq2\pi\,,\quad -\in<z<+\in \,.$$
The Poisson equation for the electrostatic potential $V_{W}^{\pm}$ associated to  a point-like charge $q$ located at $(r',\f',z') \,\epsilon\,\mathcal{M}_+$ in the wormhole manifold is
\be\lb{mw}\ba{ll}
\Delta_{({\pm};\om_0)} V_{W}^{\pm} =
\left[\frac{\pd^2}{\pd r_{\pm}^2}+\frac{1}{r_{\pm}}\frac{\pd}{\pd r_{\pm}}+
\frac{1}{\om_0^2r_{\pm}^2}\frac{\pd^2}{\pd \f^2}+\frac{\pd^2}{\pd z^2}\right] V_{W}^{\pm}
\qquad\qquad\qquad\qquad\qquad\qquad \\\\
\qquad\qquad\;\;\, =\left\{\ba{ll}
-\frac{4 \pi q}{\om_0r}\,\d(r-r')\d(\phi-\phi')\d(z-z')& \mbox{for $\mathcal{M}_+$} \, ,\\
0& \mbox{for $\mathcal{M}_{-}$}\,.
\ea\right.
\ea
\ee
Here we have used $r$  instead of $r_{\pm}$ when $\mathcal{M}_+$ or $\mathcal{M}_-$ is specified. Changing to a coordinate
$\t\equiv\om_0 \f$, the equation for the potential becomes
\be \lb{mwm}\ba{ll}
\qquad\qquad\qquad\qquad\;\;\quad\Delta_{(\pm;\om_0)} V_{W}^{\pm} =  \\
\\
\left\{\ba{ll}
\left[\frac{\pd^2}{\pd r^2}+\frac{1}{r}\frac{\pd}{\pd r}+\frac{1}{r^2}\frac{\pd^2}{\pd \t^2}+\frac{\pd^2}{\pd z^2}\right] V_W^{+}= -\frac{4 \pi q}{r}\,\d(r-r')\d(\t-\t')\d(z-z')
&\mbox{for $\mathcal{M}_+$} \, ,\\
\left[\frac{\pd^2}{\pd r^2}+\frac{1}{r}\frac{\pd}{\pd r}+\frac{1}{r^2}\frac{\pd^2}{\pd \t^2}+
\frac{\pd^2}{\pd z^2}\right] V_W^{-}= 0& \mbox{for $\mathcal{M}_-$} \, ,
\ea\right.
\ea
\ee
which corresponds to two usual potential equations (one in each manifold $\mathcal{M}_-$ and $\mathcal{M}_+$) in the space-time covered by the coordinate system $(t,r_{\pm},\t,z)$ with $r_{\pm}\geq a$ and $0\leq\t\leq2\pi\om_0$.

The solution of Eq. (\ref{mwm}) can be expanded in terms of a complete set of orthogonal functions of the coordinates $\t$ and
$z$ in the same fashion as for $V_M$ and $V_{CS}$:
\be\lb{vw}
V_{W}^{\pm}=\frac{q}{\om_0}\int \limits_{0}^{+\in}dk\cos[k(z-z')]\sum_{n=0}^{+\in}a_n \cos[\nu(\t-\t')]g_{\nu}^{\pm}(r,r')\, ,
\ee
where $\nu=n/\om_0$, $n \, \epsilon \, \mathbb{N}_0$ and
\be
a_n = \left\{ \ba{ll}
         \frac{2}{\pi} & \mbox{if $n=0$} \,,\\
         \frac{4}{\pi} & \mbox{if $n>0$}\, .
         \ea \right.
\ee
The radial functions are now separated  in three sets corresponding to three regions determined by the radial coordinate, two of them in $\mathcal{M}_+$ and the other one in $\mathcal{M}_-$, and can be written as follows:
\be\lb{radial}
g_{\nu}^{\pm}(r,r')=
\left\{\ba{ll}
g_{\nu}^{+}(r,r')=
\left\{ \ba{ll}
         \psi_{\nu}^{(1)}(kr)\psi_{\nu}^{(2)}(kr')\psi_{\nu}^{(3)}(ka) & \mbox{if $r\geq r'$}
         \\
         \psi_{\nu}^{(1)}(kr')\psi_{\nu}^{(2)}(kr)\psi_{\nu}^{(3)}(ka) & \mbox{if $r\leq r'$}
         \ea \right\}& \mbox{for $\mathcal{M}_+$}\\
         g_{\nu}^{-}(r,r')= \quad\,\psi_{\nu}^{(1)}(kr')\psi_{\nu}^{(2)}(ka)\psi_{\nu}^{(3)}(kr) & \mbox{for $\mathcal{M}_-$}
         \ea\right.
\ee
where $\psi_{\nu}^{(1)}$ and $\psi_{\nu}^{(2)}$ are linearly independent solutions of the homogeneous version of the modified Bessel equation (\ref{bessel}) of order $\nu$, satisfying the correct boundary conditions for the potential in
$\mathcal{M}_+$, and $\psi_{\nu}^{(3)}$ is another solution of homogeneous version of (\ref{bessel}), and satisfies the boundary conditions for the potential in $\mathcal{M}_-$. These functions have the general
form
\be
\psi_{\nu}^{(i)}(kr)= A_{\nu}^{(i)}I_{\nu}(kr)+B_{\nu}^{(i)}K_{\nu}(kr)\,,\quad \;\textrm{for}\; i=1,2,3.
\ee
The correct boundary conditions for the electostatic potential are listed below. Starting from the periodic conditions on $\t$, we have:
\begin{enumerate}[I.]
  \item $V_W^{\pm}(r,\t=0,z)=V_W^{\pm}(r,\t=2\pi\om_0,z)$,
  \item $\frac{\pd}{\pd \t}V_W^{\pm}(r,\t=0,z)=\frac{\pd}{\pd \t}V_W^{\pm}(r,\t=2\pi\om_0,z)$.
\begin{flushleft}
The continuity of the potential implies:
\end{flushleft}
  \item At $r=r'$: $V_W^+ \left(r\rightarrow r'^+,\t,z\right)=V_W^+ \left(r \rightarrow r'^-,\t,z\right)$,
\item At $r=a$: $V_W^+(r\rightarrow a,\t,z)=V_W^- \left(r \rightarrow a,\t,z\right)$.
\begin{flushleft}
The asymptotic behavior must fulfill:
\end{flushleft}
\item $\lim \limits_{r \rightarrow \in}V_W^+=0$,
\item $\lim \limits_{r \to \in}V_W^-=0$.
\begin{flushleft}
The requirements on the potential slope give:
\end{flushleft}
\item At $r_+=a=r_-$:
$$\frac{\pd}{\pd r_+}V_W^{+}(r\rightarrow a,\t,z)=-\frac{\pd}{\pd r_-}V_W^{-}(r\rightarrow a,\t,z),$$
\item At $r_+=r'$:
$$\frac{\pd}{\pd r} V_W^{+}(r \rightarrow r'^+,\t,z)- \frac{\pd}{\pd r}V_W^{+}(r \rightarrow r'^-,\t,z) =
-\frac{4\pi q}{r}\d(\t-\t')\d(z-z') \, .$$
\end{enumerate}
Conditions I, II, III, IV are automatically fulfilled by (\ref{vw}) and (\ref{radial}). Conditions V and VI imply that
\be
A_{\nu}^{(1)}=0=A_{\nu}^{(3)}\,,
\ee
so the other coefficients in $\psi_{\nu}^{(1)}$ and $\psi_{\nu}^{(3)}$ can be completely absorbed in
$\psi_{\nu}^{(2)}$, i.e.
\be
B_{\nu}^{(1)}=1=B_{\nu}^{(3)}.
\ee
Finally, from VII and VIII,
\be
A_{\nu}^{(2)}=\frac{1}{K_{\nu}(ka)}
\ee
and
\be
B_{\nu}^{(2)}= -\frac{1}{\left[K_{\nu}(ka)\right]^2}\left[I_{\nu}(ka)+
\frac{1}{2a\left[\frac{\pd}{\pd r}K_{\nu}(kr)\right]_{r=a}}\right].
\ee
In the neighborhood of the point charge, the electrostatic potential can always be written as
\be
V_{W}(r,\f,z)=V_M(r,\f,z)+H_{W}(r,\f,z) \, ,
\ee
where $H_{W}$ is solution of the homogeneous equation (\ref{mwm}) near the position of the charge and represents the nondivergent part of the electrostatic potential. $H_{W}$ can be seen as the sum of two contributions,
\be
H_{W}=H_0+H_{CS} \, ,
\ee
where $H_0=\left[V_{W}(r,\f,z)-V_{CS}(r,\f,z)\right]$ and $H_{CS}=\left[V_{CS}(r,\f,z)-V_{M}(r,\f,z)\right]$ (as
given in (\ref{vcsh})). The electrostatic energy of the charge at any position with $r>a$ is given by
\be\lb{ww}
W_W(r)=W_0(r)+W_{CS}(r) \, ,
\ee
where we intentionally  separated the result as the sum of the self energy corresponding to a charge in a cosmic string background  $W_{CS}(r)$  plus the extra term
\be\lb{w0}
W_0(r)=-\frac{q^2}{2\,\om_0} \int \limits_{0}^{+\in}dk\sum_{n=0}^{+\in}a_n \frac{\left[K_{\nu}(kr)\right]^2}{K_{\nu}(ka)}
\left[I_{\nu}(ka)+\frac{1}{2a\left[\frac{\pd}{\pd r}K_{\nu}(kr)\right]_{r=a}}\right]
 \,
\ee
which appears  due to the presence of a  deficit angle and a throat. This shows that the nontrivial topology of the background induces an additional contribution to the self-energy. In the particular case when $\om_0=1$, $W_{CS}=0$, so the self-energy corresponds to a cylindrical thin-shell wormhole connecting two flat geometries without a deficit angle.

The self-force deduced from (\ref{ww}) is
\be
f_{W}^r(r)=f_{0}^r(r)+f_{CS}^r(r) \, ,
\ee
where $f_{CS}^r$ is known from previous works (see \cite{lin}) and
\be
f_{0}^r(r)= \frac{q^2}{\om_0} \int \limits_{0}^{+\in}dk\sum_{n=0}^{+\in}a_n \frac{K_{\nu}(kr)}{K_{\nu}(ka)}
\left[\frac{\pd}{\pd r}K_{\nu}(kr)\right]
\left[I_{\nu}(ka)+\frac{1}{2a\left[\frac{\pd}{\pd r}K_{\nu}(kr)\right]_{r=a}}\right].
\ee
The result is best understood by plotting a numerical evaluation of the expressions above. In Figs. (1)-(4) we show the self-force as a function of the distance from the charge to the wormhole throat (normalized by the throat radius, so that $r/a=1$ corresponds to the location of the throat), for four different values of the constant $\om_0$, which is associated to the deficit angle. For the case of no deficit angle $(\om_0=1)$  the self-force is always attractive for any position of the charge; for a small deficit ($\om_0\simeq 0.9$) at both sides of the throat, the self-force is also attractive up to very large distances from the throat to the charge --but the numerical analysis shows that far enough ($r/a\sim 10^9$), the repulsive force associated to the  deficit angle becomes the dominant effect. On the other hand, when two conical geometries with larger deficit angle are connected by the throat, we can easily see that the self-force points towards the throat for small distances, while it turns to be repulsive as the distance between the throat and the charge is increased; the range such that the force points towards the throat is larger as the deficit angle decreases. Such a transition from an attractive force to a repulsive one is a  distinguishing feature of the wormhole geometry, because in the case of a cosmic string background the force is always repulsive. The attractive character and the changing behaviour of the self-force with the distance thus provide a way to detect the existence of a throat.

\begin{figure}[t!]
\centering
\includegraphics[width=10cm]{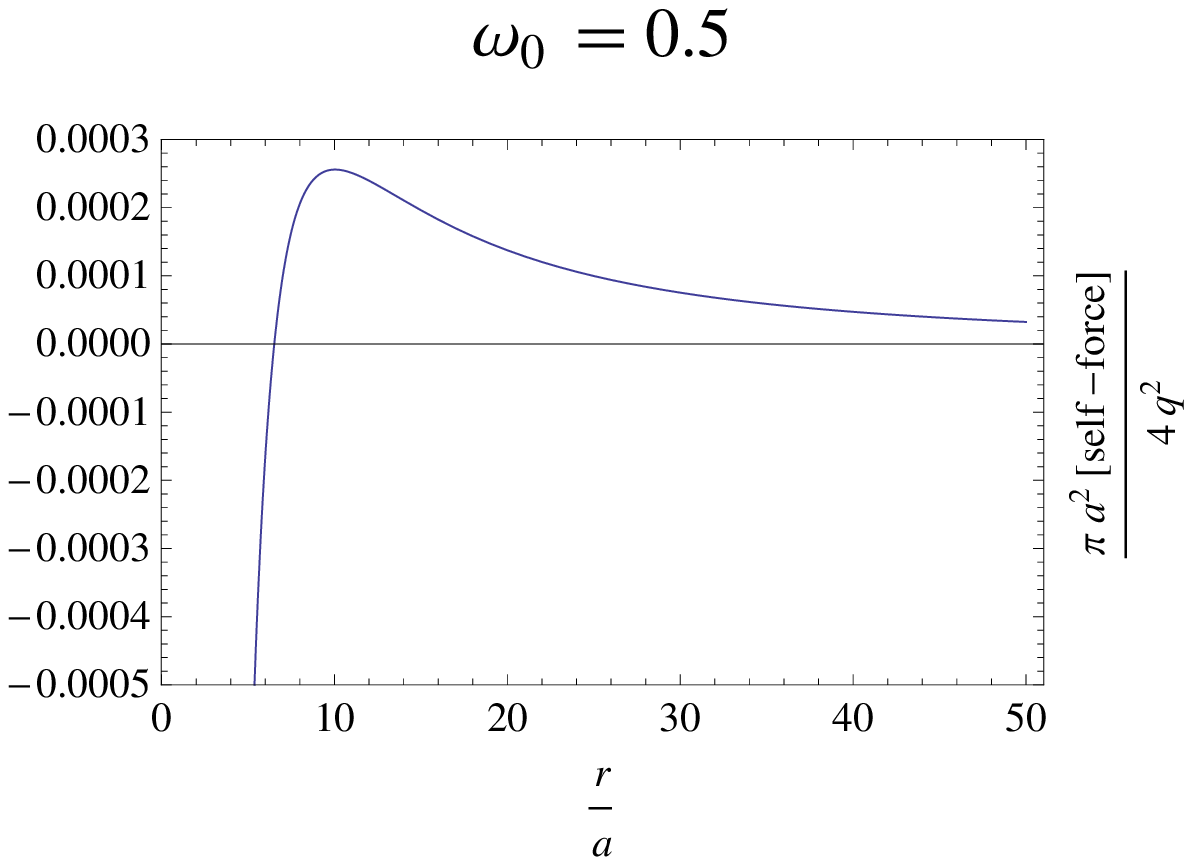}
\caption{Self-force on a charge as a function of the distance from the wormhole throat, for a large deficit angle ($\om_0=0.5$); negative values correspond to an attractive force.}
\end{figure}
\begin{figure}[t!]
\centering
\includegraphics[width=10cm]{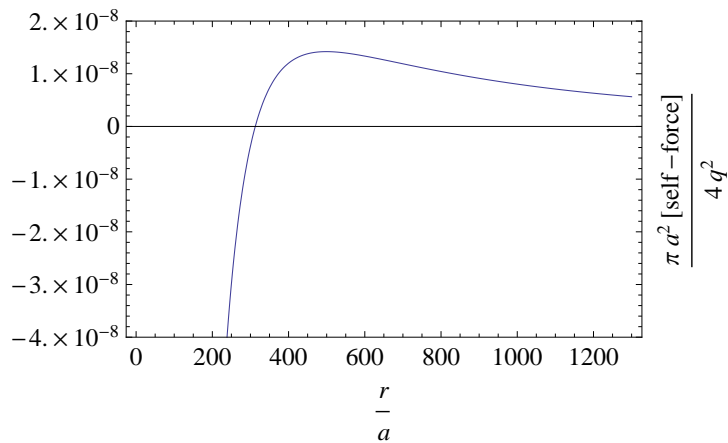}
\caption{Self-force on a charge as a function of the distance from the wormhole throat, for $\om_0=0.75$; negative values correspond to an attractive force.}
\end{figure}
\begin{figure}[t!]
\centering
\includegraphics[width=10cm]{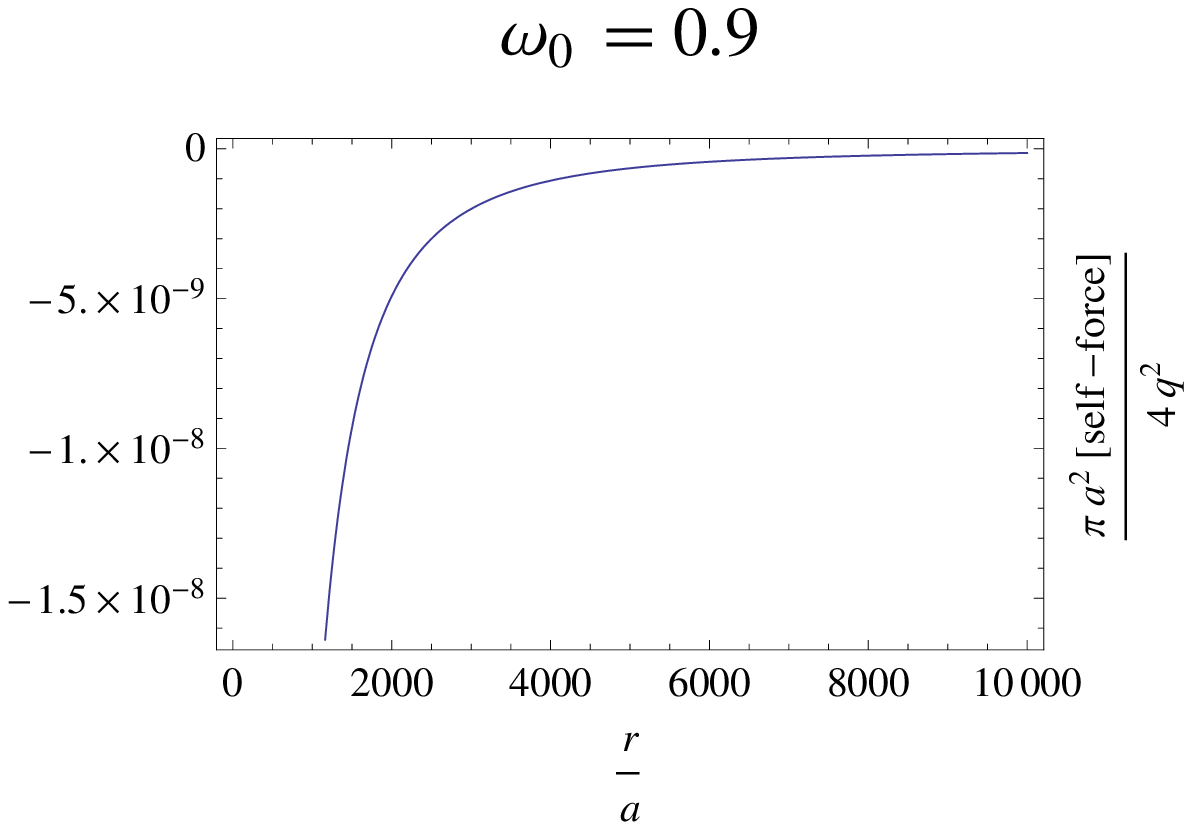}
\caption{Self-force on a charge as a function of the distance from the wormhole throat, for a small deficit angle ($\om_0=0.9$); negative values correspond to an attractive force.}
\end{figure}
\begin{figure}[t!]
\centering
\includegraphics[width=10cm]{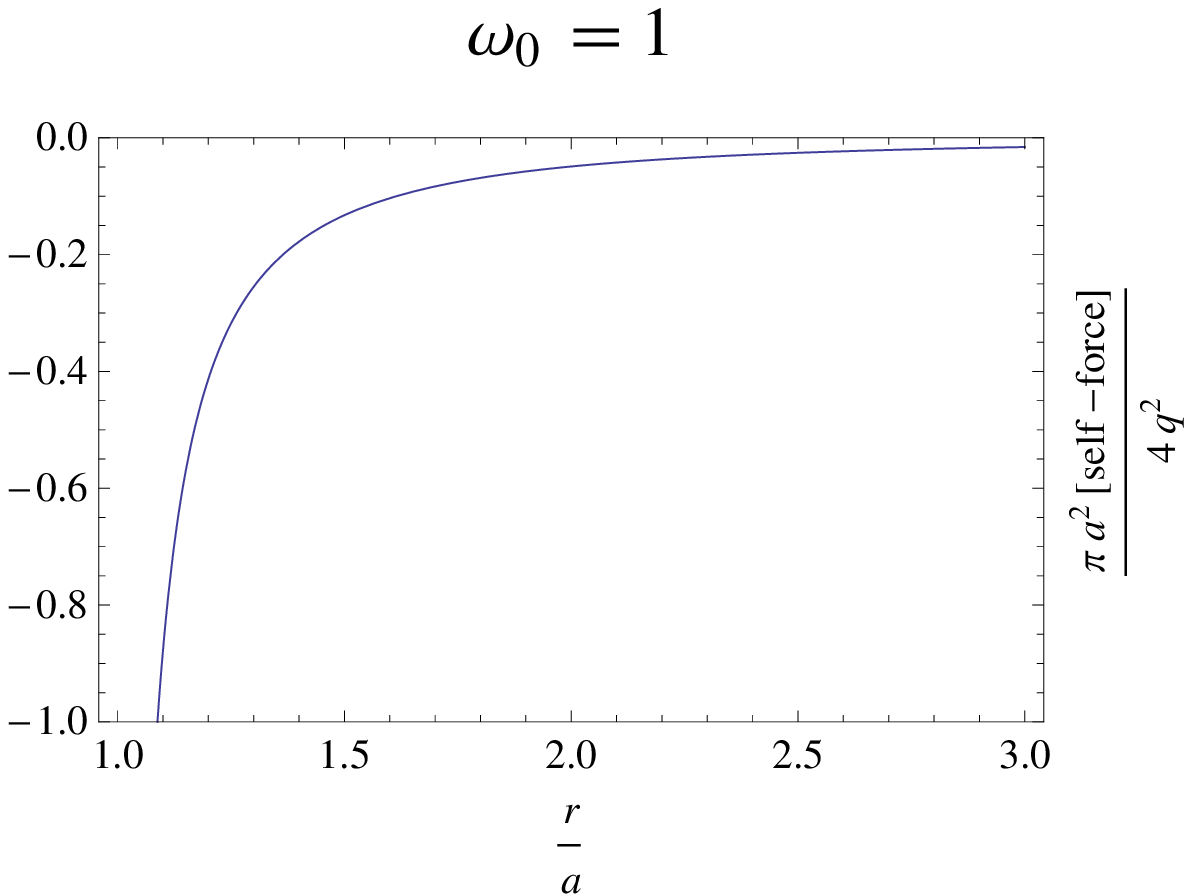}
\caption{Self-force on a charge as a function of the distance from the wormhole throat, which connects two geometries without deficit angle; negative values correspond to an attractive force.}
\end{figure}

\section{Summary}
We have evaluated the self-force on a static charged particle in the cylindrical thin-shell wormhole geometry connecting two identical flat geometries with a constant deficit angle. This wormhole can be obtained in terms of a cosmic string background by the usual cut and paste procedure, namely, by removing the regions $r<a$ from two string backgrounds and matching the resulting manifolds at $r=a$, where a throat appears. Therefore the wormhole and the cosmic string geometries are locally indistinguishable. However, their global properties are different. One one hand, it is known that the self-force in the cosmic string background always pushes the particle away from the string. On the other hand we have obtained an analytical expression in the form of a series for the self-force in the wormhole geometry and we have shown numerically that the particle is repelled from the  wormhole throat when placed beyond a certain distance from the throat, but otherwise it is attracted, as shown in figures 1 and 2. This qualitative difference is the most important result of the present work, since it shows that the study of the electrostatics of point particles in a given geometry permits to distinguish whether the topology is trivial, as for the cosmic string, or it includes a throat, as for our cylindrical wormhole. It would be of special interest to generalize these results to the Schwarzschild case, which is physically more interesting, though technically more complicated. We leave this task for a further publication.

\section*{Appendix}
The expression for the electrostatic energy and self-force for a charge in the cosmic string manifold calculated by Linet in \cite{lin} are
\be\lb{elinet}
W_{CS}(r)=\frac{L_B}{4\pi}\frac{q^2}{r}
\ee
\be\lb{linet}
f^r_{CS}(r)=\frac{L_B}{4\pi}\frac{q^2}{r^2}
\ee
respectively, where
\be
L_B=\int \limits_{0}^{+\in} \left[ \frac{\sinh(\zeta/\om_0)}{\om_0\left[\cosh(\zeta/\om_0)-1\right]}-\frac{\sinh{\zeta}}{\cosh{\zeta}-1}\right]\frac{d\zeta}
{\sinh(\zeta/2)} \, .
\ee
These results are obtained from equation (\ref{mcs}) with $0\leq\f\leq2\pi$, or equivalently from (\ref{mcsm}) where $0\leq\t\leq2\pi\om_0$ with the correct boundary conditions. For special cases, the solution for (\ref{mcsm}) that derives in (\ref{ecs}) and (\ref{fcs}) can be checked with Linet's results (\ref{elinet}) and (\ref{linet}). For example, if $\om_0=1/2$, then $0\leq\t\leq\pi$. Choosing $\t'=\pi/2$ for simplicity, the problem to solve is stated by:
\be\lb{images}
\Delta V_{CS} = -\frac{4 \pi q}{r}\,\d(r-r')\d(\t-\pi/2)\d(z-z')
\ee
and the special boundary conditions
\be\lb{cci}
 \ba{ll}
V_{CS}(r,z,0)=V_{CS}(r,z,\pi)\,,\\
\frac{\pd V_{CS}}{\pd\t}(r,z,0) = \frac{\pd V_{CS}}{\pd\t}(r,z,\pi)=0 \,.
\ea
\ee
This electrostatic problem can be solved by the method of images if the space is extended to $0\leq \t \leq 2\pi$ and another charge $q$ is located at the point $(r',\t_0=\frac{3\pi}{2},z')$. This is an equivalent problem because (\ref{images}) and (\ref{cci}) are satisfied in the region of interest. The electrostatic energy of the original charge $q$ is calculated with the potential $V_q$ generated by the image distribution. This electrostatic potential can be expanded in the same basis functions previously used in (\ref{vm}) and replacing $\t'$ by $\t_0$, so that:
\be\lb{ew1/2}
W_{CS}^{\om_0=\frac{1}{2}}(r')=\frac{q}{2}  V_{q}|_{(r=r',\t=\frac{\pi}{2},z=z')}=\frac{\,q^2}{2} \int \limits_{0}^{+\in}dk\sum_{n=0}^{+\in}a_n (-1)^n K_{n}(kr')I_{n}(kr')
\ee
The righthand side of (\ref{ew1/2}) is exactly what is obtained in (\ref{ecs}) using $\om_0=1/2$. Calculating the same energy by the usual potential expression for two point charges separated by a distance $2r'$ we have
\be\lb{ew1/2'}
W_{CS}^{\om_0=\frac{1}{2}}(r')= \frac{q}{2}  V_q(\bar{r}') = \frac{q}{2} \frac{q}{2r'}\,,
\ee
the same result obtained by Linet in (\ref{elinet}) putting $\om_0=1/2$ (so that $L_B=\pi$). The equivalence between (\ref{ew1/2}) and (\ref{ew1/2'}) implies that the self-energy calculated by both methods coincides in this particular case. Similar reasonings may be applied for $\om_0=1/m$, with $m\,\epsilon\,\mathbb{N}$, to check the result in other cases.

\end{document}